\documentclass[twocolumn]{aastex61}
\newcounter{ion}

\shorttitle{H$\alpha$ Line Profile Variations in $\beta$ Lyr}
\shortauthors{Ignace et al.}

\begin{document}
\title{A Study of H$\alpha$ Line Profile Variations in $\beta$ Lyr}


\author{Richard Ignace}
\email{ignace@etsu.edu}
\author{Sharon K.\ Gray}
\author{Macon A.\ Magno}
\author{Gary D.\ Henson}
\affil{Department of Physics and Astronomy, East Tennessee State 
University, Johnson City, TN 37663, USA }
\author{Derck Massa}
\affil{Space Science Institute, 4750 Walnut Street, Suite 205, Boulder,
CO 80301, USA}



\begin{abstract}    

We examine over 160 archival H$\alpha$ spectra from the Ritter
Observatory for the interacting binary $\beta$~Lyr obtained between
1996 and 2000.  The emission is characteristically double-peaked,
but asymmetric, and with an absorption feature
that is persistently blueshifted.  Using a set of simplifying assumptions,
phase varying emission line profiles are calculated for H$\alpha$
formed entirely in a Keplerian disk, and separately for the line
formed entirely from an off-center bipolar flow.  However, a 
dynamic spectrum of the data indicate the blueshifted feature is not 
always present, and the data are even suggestive of a drift of
the feature in velocity shift.  We explore whether a circumbinary
envelope, hot spot on the accretion disk, or accretion stream could
explain the observations.  While none are satisfactory, an accretion
stream explanation is somewhat promising.

\end{abstract}



\section{Introduction}

$\beta$ Lyr is a well-studied yet complex massive, interacting,
eclipsing binary star \citep{1980SSRv...26..349S, 1994A&A...289..411H,
2002AN....323...87H}.  Basic parameters of the system adopted for
our study are given in Table~\ref{tab1}, indicating two B-star
components in a 12.9~d period with a circular orbital separation
of $a\approx 55R_\odot$, and an accretion disk with a radius $R_D
\approx 30R_\odot$.  The deeper minimum occurs when the less massive
more luminous giant star is eclipsed.  This giant star will hereafter
be referenced as the ``Loser'' star, and the more massive component
will be the ``Gainer'' star.  The binary has been heavily studied
in the radio
\citep[e.g.,][]{1972Natur.235..270W,1978A&A....63..285J,2002A&A...391..609U},
infrared \citep[e.g.,][]{1968PASP...80...96K,1969SvA....13...83A,
1976MNRAS.174..217J,1982AJ.....87.1304Z}, optical
\citep[e.g.,][]{1996A&A...312..879H,2009ApJ...691..984S,2018arXiv180310569R,2011A&A...532A.148B},
ultraviolet
\citep[e.g.,][]{1975ApJ...198..453H,1987ApJS...65..695M,1994ApJ...421..787K},
and X-ray \citep[e.g.,][]{1976Ap&SS..42..217M,2008A&A...477L..37I}
bands.  The system has also been a frequent target for polarimetric
studies
\citep[e.g.,][]{1967ApJ...149..353A,1998AJ....115.1576H,2012ApJ...750...59L}
It is unusual for its geometrically and optically thick
disk enveloping an unseen ``primary''\footnote{Here ``primary star''
refers to the more massive component of the binary, even though the
light curve minimum in the optical occurs when the secondary star
is eclipsed.}
\citep{1963ApJ...138..342H,1974ApJ...189..319W,1991AJ....102.1156H,2008ApJ...684L..95Z}
and bipolar ``jet-like'' outflow \citep{1996A&A...312..879H,
1998AJ....115.1576H}.  Even the detection and dynamical influence of a magnetic
field has been claimed \cite[e.g.,][]{1982SvAL....8..126S,
2003A&A...405..223L}.  With such rare physical features and a
near-edge-on view, $\beta$~Lyr has been the subject of
considerable modeling, including hydrodynamic simulations
\citep[e.g.,][]{2000A&A...353.1009B,
2003ARep...47.1013N,2013ARep...57..294N}, complex stellar wind
scenarios \citep{1992A&A...254..241M}, the accretion disk modeling
\citep[e.g.,][]{2000MNRAS.319..255L,2002MNRAS.334..963L}, and the
evolutionary state of the binary \citep{2013MNRAS.432..799M}.

Our study focuses on $\beta$ Lyr's H$\alpha$ emission line.
The H$\alpha$ line of $\beta$~Lyr has proven
to be an important feature in studies of the binary.  Interferometric
studies of H$\alpha$ emission were a leading indicator of the bipolar
outflow in the binary system \citep{1996A&A...312..879H}, and
continue to be used in such approaches.  \cite{1976ApJ...210..853S}
used a set of observations of H$\alpha$ to consider relatively rapid
variations on minutes and hours (as compared to the 12.9~d period)
to interpret these effects in terms of hot spots associated with
the accretion flow onto the Gainer's disk. \cite{1978ApJ...222..627H}
focused on the asymmetry in the double-peaked H$\alpha$ line profile
shape, and the persistent blueshifted absorption feature in the
line to propose a model involving both expansion and rotation.
\cite{1993PASP..105..426H} obtained 52 spectra that were used to
create a ``dynamic spectrum'' in which spectra were binned and
grouped in 13 phase intervals to produce a gray-scale representation
of the varying emission profile.  While the authors comment that
previous studies interpreted the line shape in terms of a rotating
wind \citep{1983PASP...95..891E} or a thick disk and circumbinary
shell \citep{1973PASP...85..599B}, \cite{1993PASP..105..426H} use
their dynamic spectrum to suggest that $\beta$~Lyr bears similarity
to cataclysmic variable systems.  \cite{2009ASPC..404..297A} discuss
H$\alpha$ in terms of V/R variation, where ``V'' refers to the
blueshifted peak intensity and ``R'' refers to the redshifted one.  Their
study uses data from the University of Toledo's
Ritter Observatory (from June to
September 1999) supplemented with data from the Langkawi National
Observatory in 2007.

For our contribution to the subject, we employ a larger dataset
from the archive at the Ritter Observatory, as described in
Section~\ref{sec:obs}.  Models involving a Keplerian disk
and an off-center bipolar flow are described in Section~\ref{sec:models}.
A phase-sequenced dynamic spectrum based on the H$\alpha$ data
is given in Section~\ref{sec:disc} to interpret what appears to
be the cyclical appearance and absence, along with a drift, in the absorption
component that is seen only at blueshifted velocities.  Concluding
remarks are given in Section~\ref{sec:conc}, and an Appendix includes
details about coordinate systems used in the modeling.

\begin{table}[t]
\begin{center}
\caption{System Parameters for $\beta$ Lyr\label{tab1}}
\begin{tabular}{lcccc}
\hline \hline
Identifiers		       &  Sheliak; HD 174638; HR 7106 \\
$d$ (pc)		       &  $295\pm 15$ \\
$P_{\rm orb}$ (d)              &  12.9 \\
$\Omega_{\rm orb}$ (Hz)        &  $5.64\times 10{-6}$ \\
$a$ $(R_\odot)$                &  55 \\
$a_1$ $(R_\odot)$              &  10. \\
$a_2$ $(R_\odot)$              &  45 \\
$v_{\rm orb, 1}$ (km s$^{-1}$) &  40. \\
$v_{\rm orb, 2}$ (km s$^{-1}$) & 175 \\
$i$ $(^\circ)$                 &  86 \\
$R_D$ $(R_\odot)$	       &  30  \\ \hline
\end{tabular}
\end{center}
\end{table}

\begin{figure}
\centering
\includegraphics[width=0.95\columnwidth]{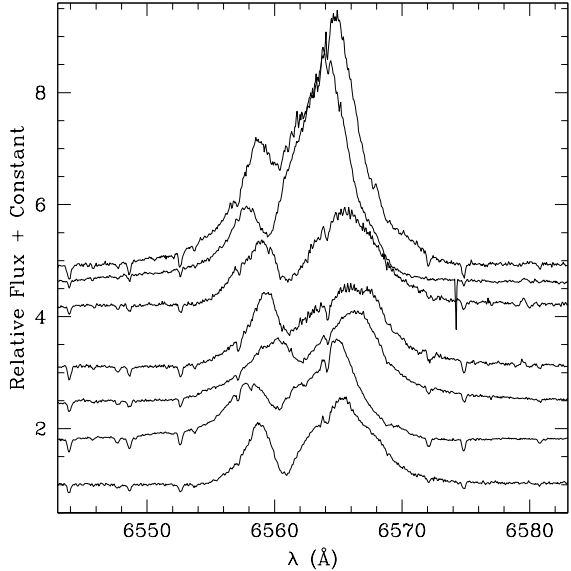}
\caption{A selection of continuum-normalized H$\alpha$ emission line
profiles from the Ritter Observatory public archive.  Each profile
has been shifted vertically for better ease of viewing.
}
\label{fig:spec} 
\end{figure}

\begin{figure}[t]
\centering
\includegraphics[width=0.95\columnwidth]{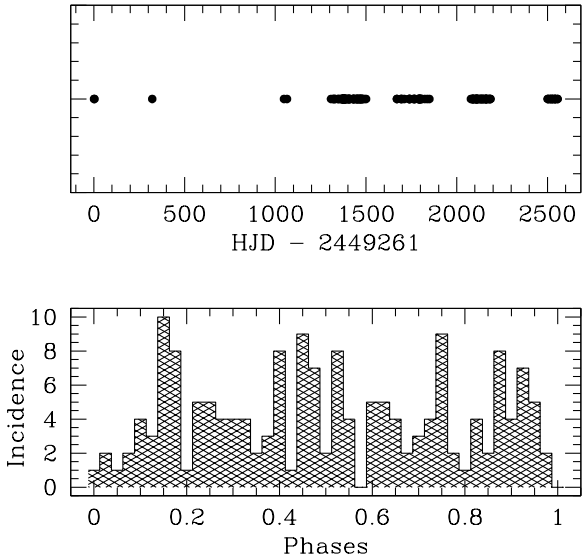}
\caption{Upper panel shows the sampling of spectra obtained
over a modified heliocentric Julian Date.  Lower panel indicates
the sampling of spectra after dates were phased on the binary
orbit (see text).  
}
\label{fig:samp} 
\end{figure}

\section{Observations of $\beta$ Lyr}	\label{sec:obs}

The Ritter Observatory\footnote{See
astro1.panet.utoledo.edu/{\textasciitilde{}}wwwphys/ritter/ritter.html for further
technical details.} operates a 1.06m reflector, employing an echelle
spectrograph with a resolving power of 26,000 at H$\alpha$
(corresponding to a velocity resolution of 11.5 km s$^{-1}$).  We
retrieved 169 CCD spectra of $\beta$~Lyr from the public archive
maintained by the Observatory.  These spectra span a period between
1993 and 2000.  For our study, we used 162 spectra taken between
1996 and 2000 (see Tab.~\ref{tab2}). Seven spectra were omitted
from this study as they were inappropriate for analysis.  Typically,
a single spectrum was obtained each night.  However, multiple spectra
(from 2 to 6) were acquired in rapid succession on eight nights in
1997, twice in 1998, and once in 2000.

The data are stored in fits files that are stacked in nine echelle orders
between the 5285 - 6595 \AA\ wavelengths.  Our study is based on
isolating the H$\alpha$ (6562.8~\AA) emission line that is in
the ninth order.  Interactive Data Language (IDL) software
routines were used to extract the H$\alpha$ data
for each of the 162 spectra.

\begin{table}[t]
\caption{Journal of Observations\label{tab2}}
\begin{center}
\begin{tabular}{clcc}
\hline\hline
Year & Months & \# of Spectra \\
\hline
1996 & August & 2  \\
1997 & April-November & 74  \\
1998 & April-October & 29  \\
1999 & June-September & 41  \\
2000 & August-September & 15  \\
\hline
\end{tabular}
\end{center}
\end{table}

With a focus on H$\alpha$, we selected a range of the wavelengths centered
on this line and performed a linear least squares fit using points
free of spectral lines to determine a continuum level.  Examples of
the continuum normalized emission profiles are shown in Figure~\ref{fig:spec}.
The profile is characteristically double-peaked but strongly asymmetric.
Line equivalent widths vary significantly, by a factor of about 3.
However, the variation is largely anticorrelated with the optical
light curve \citep[e.g.,][]{2008JSARA...2...71G}.  Peak emission in
the continuum normalized profile typically achieves a value of 2 to
3, but can rise as high as 6.  The HWHM is $\sim 200$ km
s$^{-1}$, although the wings of the line can extend to twice that.

The time of each observation was converted to heliocentric Julian date
(HJD).
The phase of each observation was calculated using the quadratic ephemeris
equation of \cite{1993A&A...279..131H}:

\begin{equation}
T = T_{\mbox{{\scriptsize0}}} + T_{\mbox{{\scriptsize1}}}\,E +
T_{\mbox{{\scriptsize2}}}\,E^{2}
	\label{eq:ephem}
\end{equation}

\noindent where $T_{0}=$HJD2408247.966,
$T_{1} =12.913780$~d represents
the binary period at $T_{0}$, and $T_{2}=3.87196 \times 10^{-6}$~d 
represents the period change.  Sampling of the archival data is
given in Figure~\ref{fig:samp}, with the upper panel showing points
at which data were taken as a modified HJD, given by subtracting
2449261.  The lower panel provides a histogram to illustrate
the sampling in orbital phase, based on equation~(\ref{eq:ephem}),
with bin widths 0.025 in phase.


\section{Line Profile Modeling}		\label{sec:models}

\begin{table}[t]
\begin{center}
\caption{Parameters for $\beta$ Lyr Stellar Components\label{tab3}}
\begin{tabular}{lcccc}
\hline\hline & Gainer & Loser \\ \hline
Sp.\ Type                    & B0V/B1V & B8III/B8II \\
$T_{\rm eff}$ (K)            & 32,000 & 13,300 \\
$M$ ($M_\odot$)              & 13 & 3 \\
$R$ ($R_\odot$)              & 6 & 13 \\
$L$ ($10^4\, L_\odot$)       & 34,000 & 4,700 \\
$v_{\rm rot}^\dag$ (km s$^{-1}$)  & 24 & 51 \\ \hline
\end{tabular}

{\small $^\dag$ Equatorial rotation speeds assuming synchronous
rotation and no oblateness.}
\end{center}
\end{table}

The $\beta$ Lyr system is a complex multi-component system consisting
at minimum of two stars, an accretion disk, a bipolar outflow that is
offset from either star, an accretion stream, and a ``hot spot'' where
the accretion stream merges with the disk.  There is even evidence for
a circumbinary envelope.  Table~\ref{tab3} lists stellar parameters
adopted for use in this study, in conjunction with system parameters
from Table~\ref{tab1}.  For Table~\ref{tab3}, the notation for stellar
properties is standard.  Note that the spectral types are debatable,
especially the Gainer star obscured by the accretion disk.  The last
row of the table gives equatorial rotation speeds of the stars assuming
synchronous rotation.  Both values are small compared to the critical
speeds of break-up, so neither component is considered rapidly rotating.
Note that in the modeling that follows, the system is so close to edge-on,
at $86^\circ$ or possibly greater (see Tab.~\ref{tab1}), that illustrative
results will assume an inclination of $i=90^\circ$ for simplicity.

\begin{figure}
\centering
\includegraphics[width=0.95\columnwidth]{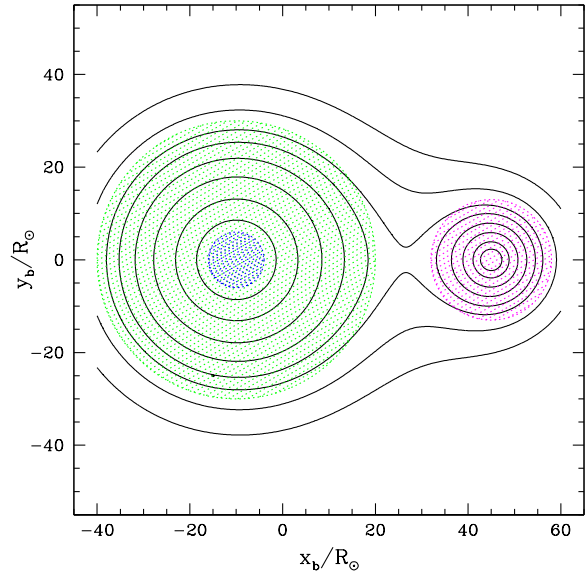}
\caption{Contours for the Roche potential in the orbital
plane.  Coordinates are in the barycenter system of $(x_{\rm b}, y_{\rm b}$,
with the barycenter at $(0,0)$.  Values are in terms of $R_\odot$.
The Gainer star is blue; the Loser is magenta; the disk is green.
All three are assumed axisymmetric.  Sizes are taken from Tabs.~\ref{tab1}
and \ref{tab3}.}
\label{fig:roche} 
\end{figure}

For illustration Figure~\ref{fig:roche} displays the Roche potential for
the system with the Gainer (blue), Loser (magenta), and disk (green)
parameters overplotted.  Isopotential values were selected to yield
somewhat regularly spaced contours in the orbital plane defined by the
barycenter (hence the ``b'' subscription; see description of coordinate
systems below).  Based on this figure, it is reasonable that the Gainer
is spherically symmetric and that much of the disk can be approximated
as axisymmetric.  The approximation of sphericity is less secure for
the Loser star, but will be adequate for our purposes.

In order to model the system, it is important to have several
coordinate systems in which to relate the various components and how
the components move with orbital phase.  Detailed descriptions for
coordinate transformations are provided in the Appendix~\ref{app:coords}.
The following highlights the principle systems and considerations.

\begin{figure}
\centering
\includegraphics[width=0.95\columnwidth]{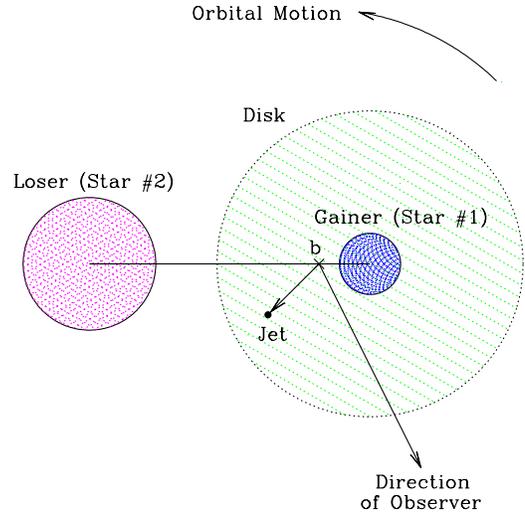}
\caption{A top-down view of the binary orbit with two stars and disk shown to
scale.  The cross marked ``b'' is the barycenter for the adopted properties
of the stars.  The dot marked ``J'' represents the origin for the jet,
as originating from somewhere in the disk.  The observer direction is
indicated for an arbitrary phase in the orbit.}
\label{fig:orb} 
\end{figure}

\begin{figure}
\centering
\includegraphics[width=0.95\columnwidth]{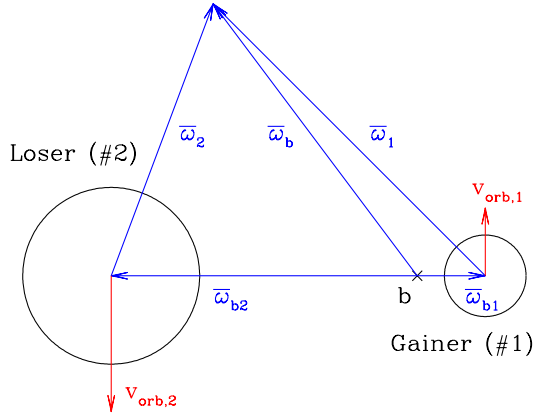}
\caption{A top-down view of the orbital plane indicating the relationship
between different coordinates, with $\varpi$ a cylindrical radius.
Also indicated are orbital velocities of the stellar components.
The sizes of the stars, their spacing, and the location of the barycenter
(indicated by ``b'') are all to scale.}
\label{fig:geom} 
\end{figure}

\begin{itemize}

\item {\em Observer Coordinates:}  Cartesian coordinates $(x,y,z)$ are
situated such that the origin is the barycenter of the binary, an
observer is along the $+z$-axis, and the $y$-axis is parallel with
the orbital angular momentum of the binary as projected in the sky.
For example, if the binary were exactly edge-on, the observer
$y$-axis would coincide with the axis of revolution for the binary,
at the barycenter.  Then the $x$-axis is defined by $\hat{x} =
\hat{y} \times \hat{z}$.  Spherical coordinates for the observer
are $(r,\theta,\alpha)$, and cylindrical coordinates are $(p,\alpha,z)$,
with $p$ an impact parameter in the plane of the sky.  Figure~\ref{fig:orb}
presents a top-down view of the orbital plane with the Gainer,
Loser, and accretion disk shown to scale (according to parameters
in Tab.~\ref{tab1}).  Here ``b'' is the barycenter.  An observer
direction is indicated at an orbital
phase of $\epsilon \approx 1/6$ (zero phase would
place the Loser directly behind the Gainer from the observer 
point of view).  Here, $\epsilon$ is the variable for phase that ranges
from 0 to 1.  A position for the center of the jet is given by way
of illustration; its exact location is not known.

\item {\em Binary Coordinates:}  With the barycenter at the origin
of the system, Cartesian coordinates for the binary are $(x_{\rm
b}, y_{\rm b}, z_{\rm b})$.  In this system $z_{\rm b}$ is the axis
of revolution for the binary orbit.  By convention primary eclipse
occurs when the Loser star is rearside of the disk.  This corresponds to
orbital phase $\epsilon = 0$.  We take the position of the Loser
to be

\begin{eqnarray}
x_{\rm b,2} = a_2\,\cos(\Omega\,t), \\
y_{\rm b,2} = a_2\,\sin(\Omega\,t), \\
z_{\rm b,2} = 0;
\end{eqnarray}

\noindent and the Gainer star to be

\begin{eqnarray}
x_{\rm b,1} = a_1\,\cos(\Omega\,t+\pi), \\
y_{\rm b,1} = a_1\,\sin(\Omega\,t+\pi), \\
z_{\rm b,1} = 0.
\end{eqnarray}

\noindent With these definitions the orbital phase tracks with
the azimuth of the Loser star, $\varphi_{\rm b,2}$, 
$\epsilon = \varphi_{\rm b,2}(t)/2\pi$.

Spherical coordinates in the barycenter system are $(r_{\rm b},
\vartheta_{\rm b}, \varphi_{\rm b})$, and cylindrical coordinates
are $(\varpi_{\rm b}, \varphi_{\rm b}, z_{\rm b})$.

\item {\em Stellar Coordinates:}  Notation for the star coordinates follow
those for the barycenter, but with subscripts of ``1'' for the Gainer
and ``2'' for the Loser.  The orientation of the Cartesian coordinates
for the three systems have all of the $x$, $y$, and $z$ axes similarly
oriented.  For example, $+z_{\rm b}$, $+z_1$, and $+z_2$ are all
parallel.  If $x_{\rm b,1} = -a_1$ and $x_{\rm b,2} = +a_2$, then
the $x_1$-coordinate for the barycenter is $+a_1$, and 
the $x_2$-coordinate is $-a_2$.  Figure~\ref{fig:geom} provides
an illustration of the relative positioning of cylindrical
radii for the barycenter, Gainer, and Loser systems.  The 
three vectors in the upward direction in the figure are for
an arbitrary point in the orbital plane.  Also indicated 
are the instantaneous orbital velocity vectors of the stars
(values given in Tab.~\ref{tab1}).

Note that the accretion disk is centered on the Gainer star, and so
the coordinates for the disk are conveniently defined in the Gainer
star system.

\item {\em Jet Coordinates:}  The last system is the jet itself.  When
calculating line profile shapes for the jet component, it will prove
convenient to define properties such as density and velocity in the
jet system, and then use coordinate transformations to the observer
system.

The jet is taken to be a biconical outflow, where the vertex is in
the orbital plane, with symmetry axis normal to that plane.  Cartesian
coordinates in the jet system are $(x_J, y_J, z_J)$, with $z_{\rm
J}$ the symmetry axis of the cone.  The $x$-axis for the jet
corresponds to the outward cylindrical radial from the Gainer star,
which in unit vectors is $\hat{x}_J = +\hat{\varpi}_{\rm 2,J}$.
Spherical and cylindrical coordinates in the jet system follow that
of the barycenter and the two stars.

\end{itemize}

\subsection{Line Profiles from the Disk}

\subsubsection{Theory}

An axisymmetric, rotating circumstellar disk can generally be expected
to produce emission line profiles that are double-peaked in shape
\citep[e.g.,][]{1986MNRAS.218..761H}.  The objective of our modeling is to obtain
qualitative features for H$\alpha$ emission produced in the disk, and
to simulate
how the profile shape varies with orbital phase.  This involves modeling
not only the line formation throughout the disk, but also attenuation
by the continuous opacity of the thick disk, plus the effect of
eclipse by the Loser star.  Loosely following \cite{1983ApJ...274..380R},
we use Sobolev theory \citep{1960mes..book.....S} for an edge-on disk
in Keplerian rotation.  

The disk is taken to be axisymmetric with a density profile described
most conveniently in the stellar coordinates of the Gainer, as given by

\begin{equation}
\rho(\varpi_1,z_1) = \rho_0 \, e^{-|z_{\rm 1}|/H_D}\,e^{-\varpi_1^2/R_D^2},
\end{equation}

\noindent where $\rho_0$ is a constant that is set by the total
mass in the disk.  Estimates for the dimensions of the disk are
$R_D \approx 5 R_1$ and $H_D \approx R_1$.  The density has
an exponential decline vertically, but a Gaussian decline in cylindrical
radius.  The latter serves to rather rapidly truncate the radial extent
of the disk.

The Gainer star has a critical speed of rotation at its equator of

\begin{equation}
v_{\rm c,1} = \sqrt{\frac{GM_1}{R_1}} \approx 640~{\rm km~s^{-1}}.
\end{equation}

\noindent Hereafter, it is convenient to normalize lengths to
the radius of the Gainer.
The Keplerian rotation in the disk (taken as uniform with
height in the slab) is

\begin{equation}
v_K(\varpi) = v_{\rm c,1}\,\sqrt{\frac{1}{\varpi}}.
\end{equation}

The Doppler shift with respect to a distant observer is

\begin{equation}
v_{\rm z} = \vec{v}_K\cdot\hat{z} + \vec{v}_{\rm orb}\cdot\hat{z},
\end{equation}

\noindent where the second term represents an additional velocity shift
owing to the participation of the disk in the binary orbit about the
barycenter.  The disk is assumed stationary in the co-rotating frame.
As a result, in the observer frame, locations in the disk have a Doppler
shift arising from solid body rotation about the barycenter.  The 
observed velocity shift becomes

\begin{equation}
v_{\rm z} = v_{\rm c,1}\,\frac{\sin\varphi_1}{\sqrt{\varpi}}
	+ \Omega\,R_1\,\varpi_{\rm b}\,\sin\varphi_{\rm b}.
\end{equation}

The critical speed of rotation for the Gainer is about 640 km~s$^{-1}$.
The outer edge of the disk rotates with a Keplerian speed of 290
km~s$^{-1}$.  The maximum solid body rotation speed in the disk,
relative to the barycenter, is 160 km~s$^{-1}$.  While 160 km~s$^{-1}$
is non-trivial compared to the slowest Keplerian speed in the disk,
much of the disk has a solid body rotation that is considerably
smaller than the Keplerian speed.  As a result, we model disk line
profiles based on Sobolev theory for a disk with no solid body
rotation, and then include a shift of the resulting profile as a
whole with an amplitude that mimics that of the Gainer star's radial
velocity shift, with respect to the barycenter.  The consequence
of this simplification is that, in the absence of an eclipse by the
Loser, our disk line profiles are left-right symmetric about a
bisector.  With solid body rotation, we would expect a double-peaked
line shape that is slightly asymmetric in a phase-dependent way,
owing to how the solid body rotation term generally produces a
differential distortion of the isovelocity zones between the blue
and redshifted sides.  (Note that the phase-averaged line profile
would still be symmetric about line center.)

Following \cite{1983ApJ...274..380R}, the Sobolev optical depth is given by

\begin{equation}
\tau_S = T_L\,\left(\frac{\rho}{\rho_0}\right)^2\,\frac{\varpi_1}{2\,| w_{\rm z}
	|\,\sqrt{1-\varpi_1\,w_{\rm z}^2}},
\end{equation}

\noindent where $T_L$ is a free parameter for the scaling of the line
optical depth, $\rho_0$ is the peak density in the disk, the square
of density is a scaling appropriate for modeling H$\alpha$ as
a recombination line.  The normalized line-of-sight Doppler velocity shift
in the profile is $w_{\rm z}$ for emission arising in
the disk.  It is normalized to the critical speed of rotation for the
Gainer star and is given by

\begin{equation}
w_{\rm z} = v_{\varphi,1}/v_{\rm c,1} = \frac{\sin\varphi_1}{\sqrt{\varpi_1}},
\end{equation}

\noindent where $v_{\varphi,1}$ is the Keplerian azimuthal speed is
the disk.  Consequently, $w_{\rm z}$ ranges from $-1$ to $+1$ for this case.

An isovelocity zone is a volume in the disk for which $w_{\rm
z}$ is a constant.  An example of such zones can be found in
\cite{2010ApJ...725.1040I} (Fig.~2 of that paper).  The contour for
constant $w_{\rm z}$ is a dipole-like loop with inner radius $\varpi=1$ at
the surface of the Gainer star, and outer radius $\varpi_{\rm max}(w_{\rm
z}) = 1/w_{\rm z}^2$.

An unocculted edge-on disk is for simplicity taken to produce a specific flux
of continuum emission given by:

\begin{equation}
f_D = \frac{2H_D\cdot 2R_D}{d^2} \cdot S_D\,\int_0^\infty\,
	\left(1-e^{-\tau_D}\right)\,dp,
\end{equation}

\noindent where $d$ is distance to the star, $\tau_D$ is the disk
optical depth, $2H_D$ is from the vertical integral through the disk,
and $S_D$ is the source function in the disk, here taken as a constant.
Another factor of two appears because the integration limits are for
just half of the disk.  Note for an edge-on disk with $\tau_D \gg 0$,
the flux becomes $f_D \approx (\sqrt{\pi}\, H_D\,R_D/d^2)\,S_D$.

The disk optical depth is expressed as

\begin{equation}
\tau_D = 2T_C\,e^{-p^2/R_D^2}.
\end{equation}

\noindent Here $2T_C$ is the total optical depth along a diameter
through the disk, as if the Gainer star were not present.  For
convenience the product of opacity and density as integrated along
a chord in the orbital plane is here assumed to yield a Gaussian
dependence with the observer's impact parameter, $p$.  

The specific flux of line emission at $w_{\rm z}$ is given by

\begin{eqnarray}
f_L(w_{\rm z}) & = & 2H_D\,R_1\,\int_1^{\varpi_0(w_{\rm z})}\,
	S_L \, \left( 1- e^{-\tau_S}\right) \times \nonumber \\
     & & \frac{3}{2}\,w_{\rm z}\,
	\varpi_1\,\Lambda(\varpi_1,w_{\rm z})\,d\varpi_1,
\end{eqnarray}

\noindent where $S_L$ is the line source function,
and the appearance of $2H_D$ indicates that the result for
the midplane of the disk applies for every height in the disk slab.

The factor $\Lambda$ accounts for the attenuation of line photons by
the continuous opacity of the optically thick disk.  The isovelocity
zones for an axisymmetric, rotating disk are left-right symmetric.
The disk rotates counter-clockwise as seen from above, so blueshifts
arise on the left side for an observer, and redshifts on the right side.
A single isovelocity contour is a loop, and therefore has a front-side
branch and back-side branch.  The extinction of line photons on the near
branch is less than for the back.

\begin{figure}[t]
\centering
\includegraphics[width=0.95\columnwidth]{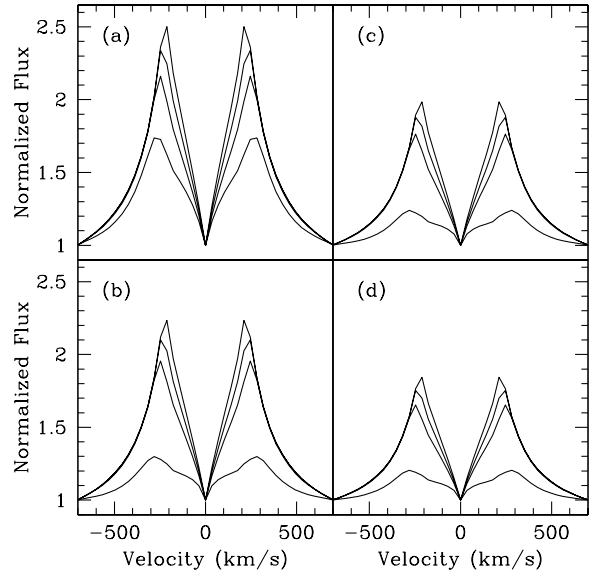}
\caption{A series of model line profiles formed in the Keplerian accretion
disk.  Each panel has 4 line profiles, with $T_L = 0.3, 1.0, 3.0,$
and 10.0.  The different panels are for different optical depths
in the disk, with $T_C = 0.3, 1.0, 3.0,$ and 10.0 for panels (a)--(d),
respectively.}
\label{fig:disklines} 
\end{figure}

The factor $\Lambda$ is given by

\begin{equation}
\Lambda = e^{-\tau_{D,A}} \, e^{-\tau_S} + e^{-\tau_{D,B}}.
\end{equation}

\noindent where $A$ and $B$ are respectively for the near and far
branches, and the additional factor relating to the Sobolev optical depth
is a recognition that line photons produced in the far branch are also
attenuated by the line opacity in the near branch \citep{1978ApJ...219..654R}.
Note that for an optically thin line and optically thin disk, $\Lambda
\approx 2$.  This is because the flux integral is for just the near
branch, but both front and rear
contribute to the line at normalized velocity shift
$w_{\rm z}$.  If the disk is extremely thick, $\Lambda \rightarrow 0$.

\begin{figure}[t]
\centering
\includegraphics[width=0.95\columnwidth]{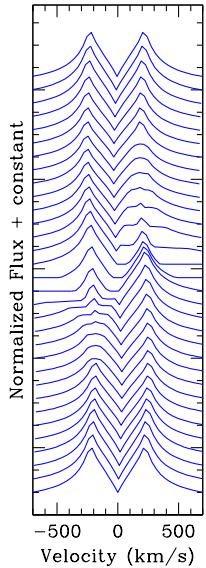}
\caption{A series of disk line profiles with orbital phase
for $T_L=10.0$ and $T_C=3.0$.  The profiles are vertically offset
by a constant for ease of viewing, with phase $\epsilon=0$ at bottom
and $\epsilon=1$ at top.  There are 33 profiles shown.  The phase interval
between profiles is $\Delta \epsilon \approx 0.03$.  The distortion to
the profiles around $\epsilon$ of 0.5 arises from an eclipse of the disk
by the Loser, which first affects the blueshifted side, then the redshifted
side.}
\label{fig:diskphase} 
\end{figure}

\subsubsection{Results}

Given the prescribed geometry of the disk, and treating the system
as viewed edge-on, the main free parameters of the model are
$T_L$ to control the line strength, $T_C$ to control the
severity of disk extinction, $S_L$ for the line source function,
and $S_D$ for the disk source function.  
In addition, we assume that at
H$\alpha$, the line source function scales as
a blackbody in the Rayleigh-Jeans (RJ) limit at isothermal temperature 
${\cal T}_L=20,000$~K.  For $S_D$, the disk is taken to emit like
a blackbody of temperature ${\cal T}_D =9,000$~K, also in the RJ limit.

Figure~\ref{fig:disklines} illustrates profile shapes for a range of
different optical depth combinations.  Panels (a) through (d) are for
$T_C = 0.3, 1.0, 3.0,$ and 10.0, respectively.  In each panel the four
line profiles are for $T_L = 0.3, 1.0, 3.0,$ and 10.0, with stronger lines
resulting for larger $T_L$ values.  All the line profiles are displayed
as continuum normalized.  In these examples, the Loser star is at orbital
phase 0.0, so there is no eclipse of either the Loser or the disk, to
influence either the profile shape or the continuum normalization.  The
normalization takes into account flux from the Loser star and the disk,
including generally when either the Loser or the disk are in eclipse.

The profiles are mainly illustrative.  They demonstrate that for an
axisymmetric disk, a line shape that is double peaked and symmetric about
line center results.  When the line is optically thick, a distinct
wedge-shaped central depression results.  When the line is more
optically thin, the central depression has more structure.  
\cite{1983ApJ...274..380R} considered rotating rings and found
a range of profile morphologies from ``M'' shapes for thick lines
to ``U'' shapes for thin ones.  Our construction, based on
their approach, is a series of nested rings, each having its own
optical depth, but modified according to
absorption by continuous opacity of the disk.

A few interesting points.  (1) The peaks are less separated as $T_L$
increases.  (2) Peak emission is overall depressed as $T_C$ increases.
Note that none of these examples achieve peak values as large
as the observed maximum.  With a disk that is essentially truncated, 
increasing $T_L$
has limits for increasing the line emission.  At some point the
only way to obtain a stronger line is to increase $S_L$.  (3) While
the line peaks extend to $\approx \pm 300$ km s$^{-1}$ (somewhat
higher values than observed), the emission wings can go to rather
large values, beyond $\pm 500$ km s$^{-1}$.  The highest speed line
emission originates from close to the star, where Keplerian rotation
is greatest.  The maximum Keplerian speed is 640 km
s$^{-1}$ at $\varpi_1=1$ (i.e., radius of the Gainer star).

Figure~\ref{fig:diskphase} shows a series of disk profiles as a function of
orbital phase.  Each profile is shifted vertically for ease of viewing.
The profile at bottom is for phase $\epsilon = 0.0$; the topmost is for
$\epsilon = 1.0$.  The development of asymmetry at middle phases, around
$\epsilon=0.5$, arises when the Loser eclipses the disk.  The eclipse
first affects the blueshifted velocities, and then migrates across to
redshifted ones.  In these examples, a significant fraction of the line
emission can be blocked, but the continuum normalization also evolves,
because the disk is being eclipsed.

\subsection{Line Profiles from the Jet}

\subsubsection{Theory}

To model emission line profiles from a jet, Sobolev theory is again
employed; however, the geometry and velocity field are quite different
from the disk case.  The jet is approximated as a biconical flow, with
center situated in the disk.  We again assume that the system is
viewed edge-on.  Next we treat the biconical flow as being a portion
of a spherical wind.  In other words the velocity and density in
the flow are functions only of radius from the vertex of the cone,
$r_J$:  the density producing H$\alpha$ is nonzero within the biconical
of half-opening angle $\beta_J$, and zero outside of it.

For the velocity field, the flow is 
assumed to expand homologously, with no rotation, with

\begin{equation}
v(r_J) = v_0 \, \left(\frac{r_J}{R_1}\right),
\end{equation}

\noindent where $R_1$ is used to normalize the radius, and $v_0$ is a scale
parameter for the velocity.  Mass continuity provides the density
in the jet, with

\begin{equation}
\rho(r_J) = \frac{\dot{M}}{4\pi\,r_J^2\,v(r_J)}.
\end{equation}

\noindent The biconical geometry implies that the density diverges
at the vertex.  This does not present a practical problem, since
the vertex also has vanishing volume for purposes of calculating the line
flux.

The isovelocity zones for homologous expansion are well-known to
be planes oriented to be normal to the viewer sightline
\citep[e.g.,][]{1978stat.book.....M}.  Interestingly, the shapes defined
by the intersection of such planes with the biconical flow geometry
leads to conic sections, with the type depending generally on how the viewing
inclination compares to the cone opening angle.  For edge-on the emitting
surfaces are parabolae, with a gradient in surface brightness that drops
quickly with height from the orbital plane.

The Doppler shift for a point in the jet as seen edge-on is

\begin{eqnarray}
v_{\rm z} & = & -\hat{z} \cdot \vec{v} + \Omega\,R_1\,\sin(\varphi_{\rm b}) \\
          & = & -v_0\,\frac{r_J\,\cos\theta}{R_1} 
	+ \Omega\,R_1\,\varpi_{\rm b}\,\sin(\varphi_{\rm b}) \\
          & = & -(v_0/R_1)\,z + \Omega\,y + x_{\rm bJ},
\end{eqnarray}

\noindent where $y$ and $z$ are observer coordinates, and

\begin{equation}
x_{\rm bJ} = \varpi_{\rm bJ}\,\cos (\Omega t - \varphi_{\rm bJ}).
\end{equation}

\noindent Note that if the jet were not participating in orbital motion,
the isovelocity zones would be planes normal to the observer sightline
$\hat{z}$.  The above expression reveals that for the biconical jet
with homologous expansion, in conjunction with the solid body
rotation (i.e., the biconical geometry is defined for the rotating
frame), the isovelocity zones are still planar, but with a normal
that is rotated with respect to $\hat{z}$.  In the observer
coordinates, the planes have normals, $\hat{n}$, with components:

\begin{eqnarray}
n_{\rm x} & = & \frac{\Omega\,R_1}{\sqrt{v_0^2+\Omega^2R_1^2}}, \\
n_{\rm z} & = & \frac{v_0}{\sqrt{v_0^2+\Omega^2R_1^2}}.
\end{eqnarray}

\noindent The deviation of the isovelocity planes from having normals
of the observer sightline $\hat{z}$ 
is given by $\cos \psi = \hat{z}\cdot\hat{n} = n_{\rm z}$.
The product $\Omega R_1 = 24$ km~s$^{-1}$ is the speed of synchronous
rotation at the equator of the Gainer.  By contrast, one expects that
$v_0\sin\beta_J$ to approximately set the half-width of an
emission line formed in the jet.  For $\beta_J \sim 60^\circ$
and the observed half-width around 300 km~s$^{-1}$, $v_0$ must be around
350 km~s$^{-1}$, demonstrating that the influence of solid body rotation
is mild.  

The Sobolev optical depth for a radial outflow in homologous expansion is

\begin{equation}
\tau_S = T_J\,\left(\frac{\rho}{\rho_0}\right)^2\,q(r),
\end{equation}

\noindent where $T_J$ is a line optical depth scale for the 
jet, and $q$ is a function of radius that can subsume
modifications to the recombination rate and other factors that
alter the optical depth at H$\alpha$.  Here, we use a power law, with
$q \propto r^\delta$ as a way to make $\tau_S$ vary more or less steeply
than $\rho^2$ with radius.  The flux of line emission becomes

\begin{equation}
f_\nu(w_{\rm z}) = \frac{1}{d^2}\,\int_{\varpi(w_{\rm z})}^\infty
	\, S_J\,\left(1-e^{-\tau_S}\right)\,4\alpha_0(r_J,v_{\rm z})
	\, r_J\,dr_J,
\end{equation}

\noindent where $S_J$ is the line source function for the jet.  As
with the disk, the jet is taken to be a source of blackbody radiation
in the RJ-limit with temperature ${\cal T}_J = 20,000$~K.  For a
spherical wind, $\alpha_0=\pi/2$.  For an edge-on biconical jet,
$\alpha_0$ is the half opening angle of the parabolic isovelocity
zone with observer impact parameter $p$.  The factor of $4\alpha_0$
spans the full width of the upper and lower parabolae.  Its value
is given by the expression

\begin{equation}
\sin\alpha_0 = \sin\beta_J\,\sqrt{1-\frac{p^2_{\rm min}(v_{\rm z})}{p^2}},
\end{equation}

\noindent where $p_{\rm min}$ is the impact parameter at the apex of
the parabola as given by

\begin{equation}
p_{\rm min} = | z | / \tan \beta_J,
\end{equation}

\noindent and $p = \sqrt{r_J^2 - z^2}$, for the isovelocity plane for
which $z=R_1\,v_{\rm z}/v_0$.

The effect of orbital motion is included for the line profile
as follows.  Since we have
argued that modification to the isovelocity zones is rather small,
we approximate the effect of orbital motion as a bulk velocity shift of
the entire line shape by the amount:

\begin{equation}
\Delta v_{\rm z} = +\Omega\,\varpi_{\rm bJ}\,\sin(\Omega t+\varphi_{\rm bJ}),
\end{equation}

\noindent where the product $\Omega\,\varpi_{\rm bJ}$ is the solid body
rotation speed of the jet origin in the orbital plane, and 
$\varphi_{\rm bJ}$ is the azimuthal offset of the jet origin, about
the barycenter, from the line-of-centers between the two stars.

The examples in the next section place the jet origin on the
line-of-centers at the edge of disk, for which $\varphi_{\rm
bJ}=0^\circ$ and $\varpi_{\rm bJ} \approx 3.3$, so that
$\Omega\,\varpi_{\rm bJ} = 77$ km s$^{-1}$.  We choose $\delta=1$
giving $q \propto r$, and $v_0 = 150$ km s$^{-1}$.

\subsubsection{Results}

\begin{figure}
\centering
\includegraphics[width=0.95\columnwidth]{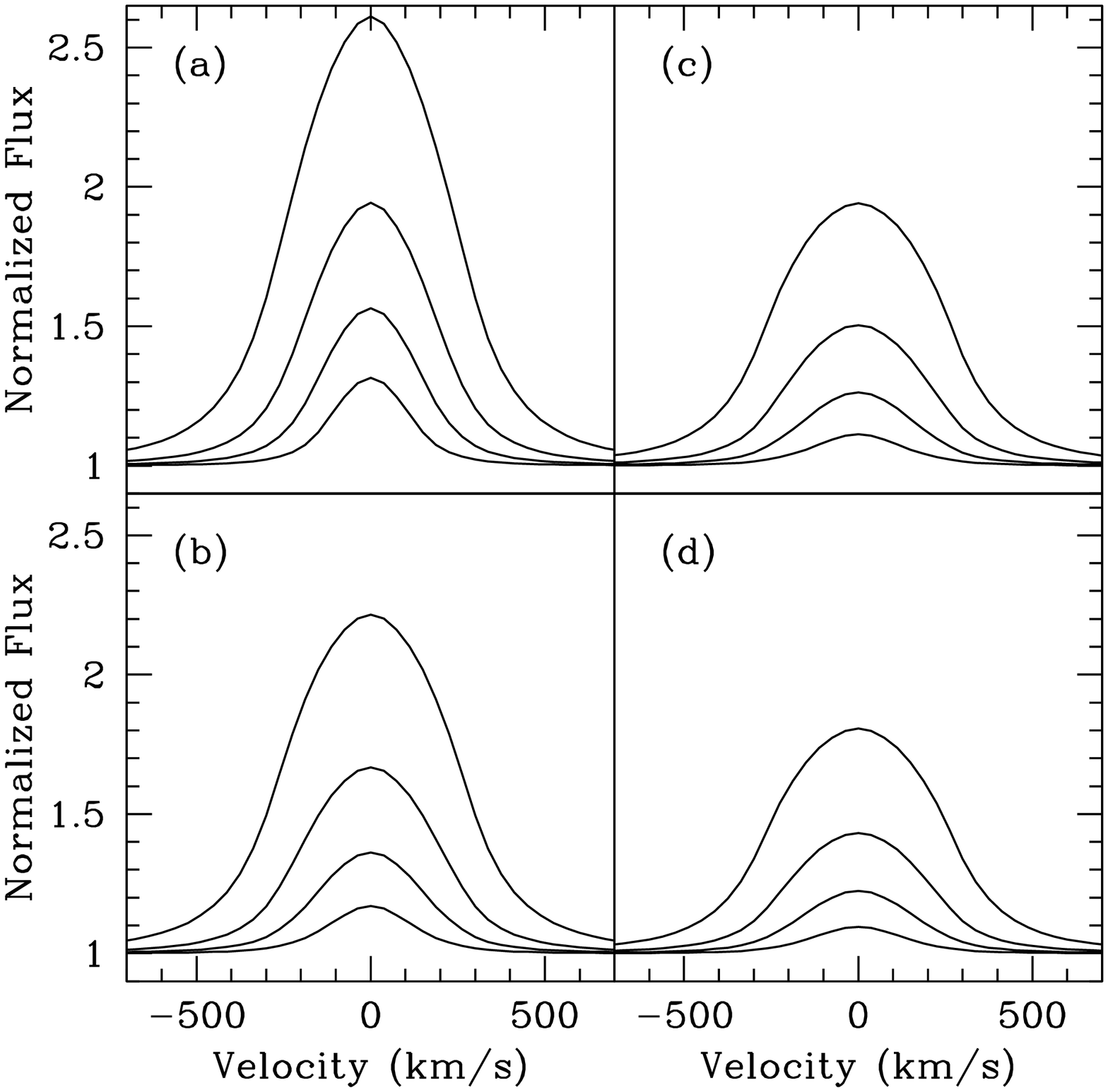}
\caption{A series of model line profiles formed in the biconical
jet with homologous expansion.
Each panel has 4 line profiles, with $T_L = 3.0, 10.0, 30.0,$
and 100.0.  The different panels are for different optical depths
in the disk, with $T_C = 0.3, 1.0, 3.0,$ and 10.0 for panels (a)--(d),
respectively.}
\label{fig:jetlines} 
\end{figure}

Figure~\ref{fig:jetlines} displays a sample of model line profiles
for the biconical jet following the format for the disk case in
Figure~\ref{fig:disklines}, and also evaluated
at a phase $\epsilon=0.0$.  The optical depth scale $T_C = 0.3,
1.0, 3.0,$ and 10.0 for panels (a)--(d), as in the disk case.  However,
the four profiles are, with increasing line strength, $T_L =
3.0, 10.0, 30.0,$ and 100.0, which is different from the disk case.

Note that the jet model is distinct from the disk model in two particular
ways.  First, the disk is essentially truncated.  In the orbital plane,
its extent is limited by the presence of a binary companion.  This is
different from the expansive accretion disks that can form around pre-main
sequence stars \citep[e.g.,][]{2009ApJ...701..260I}.  Vertically,
the disk is limited by gravity and gas pressure-support.  
All else being equal, the line equivalent width for the disk
grows slowly once the line becomes quite optically thick.  In this regime
the line flux scales as a product of the projected area of the source 
and the source function, and the area is limited.  On the other hand,
although the jet is conical in angular extent, it is not limited
in radial extent.  Although unphysical, the line equivalent is 
formally unbounded in the model as $T_L$ is increased.

Second, the jet does not easily produce a traditional P~Cygni line
profile \citep[e.g.,][]{1999isw..book.....L} because the jet is offset from
the Gainer star.  Typically, a jet seen at arbitrary
inclination angle tends to produce a double-peaked line profile.
If a jet originates from a star, then the outflowing portion directed
toward the observer will also lie in front of the star, and so can
produce the standard blueshifted absorption trough characteristic
of a P~Cygni line\footnote{Even for a recombination line that may
not display a net absorption because the emission fills it in, one
generally expects the blueshifted absorption to cause asymmetry in
the line shape.}.  For $\beta$~Lyr, the system is very close to
edge-on.  This means the line tends to be single-peaked instead of
double-peaked.  Being offset from the star means that when the jet
is forward of the Gainer, then if there were no disk, and if the
system were edge-on, the absorption trough would be symmetric about
line center, and a standard P~Cygni line shape cannot result.
Consequently, the model emission lines for the jet can grow to high
peak values but are always single-peaked.

\begin{figure}
\centering
\includegraphics[width=0.95\columnwidth]{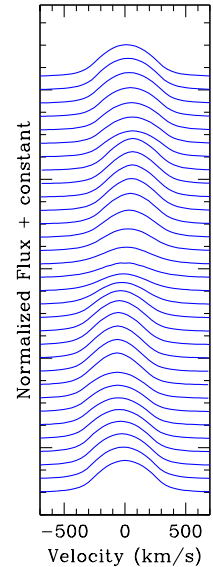}
\caption{A series of jet line profiles with orbital phase
for $T_L=100.0$ and $T_C=3.0$, following the style of Fig.~\ref{fig:diskphase}.
The depression of
the profiles around $\epsilon$ of 0.5 arises from an eclipse of the jet
by the Loser.  For this example, the jet is at the edge of the disk,
on the line of centers for the two stars.
}
\label{fig:jetphase} 
\end{figure}

Figure~\ref{fig:jetphase} for jet lines with phase variations is
similar to Figure~\ref{fig:diskphase} for the disk case.
Each profile is for a different phase of the orbit, with $\epsilon=0.0$
at bottom, and $\epsilon=1.0$ at top.  The slight modulation in
location of the peak is the effect of orbital motion.  The
depression of the line emission at mid-phases is from the eclipse
of the jet by the Loser star.  Whereas the isovelocity zones for
the disk were left-right symmetric so that the eclipse led to line
asymmetry, the isovelocity zones for the jet are back-front symmetric
in our model, and so the line profile remains symmetric about peak
emission at all times.

\begin{figure*}[t]
\centering
\includegraphics[width=1.95\columnwidth]{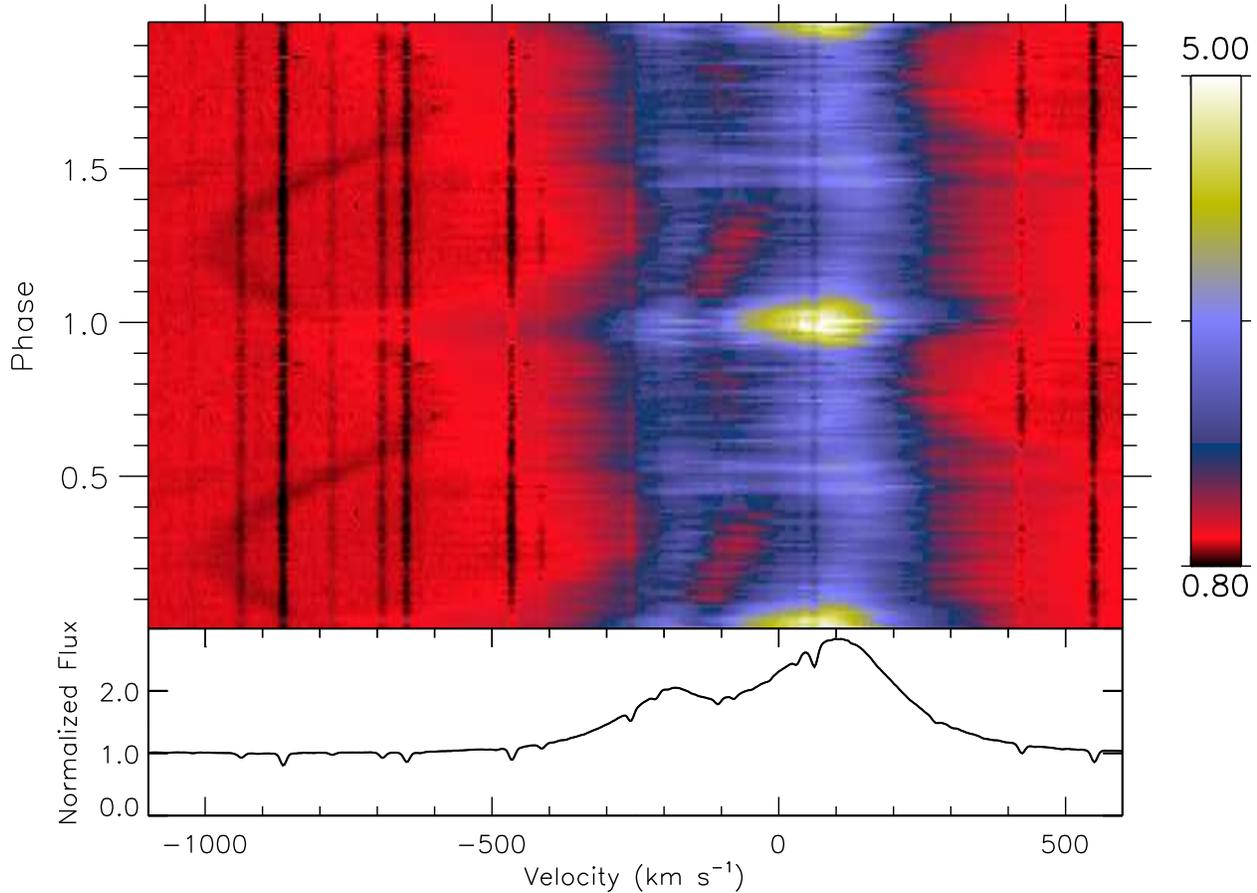}
\caption{A dynamic spectrum for the phased continuum
normalized spectra in the vicinity of H$\alpha$.
Bottom shows the average profile shape.  A color scale for brightness
is indicated at right.  The data in phase is repeated for ease of
viewing.  The sinusoid absorption feature in black centered near
$-800$ km s$^{-1}$ is associated with the Loser star.
}
\label{fig:dynamic} 
\end{figure*}

\section{Discussion}	\label{sec:disc}

Figure~\ref{fig:dynamic} shows a dynamic spectrum for the set of 
continuum-normalized H$\alpha$ data.
Spectra in the vicinity of H$\alpha$ are displayed in phase order along
the vertical, with velocity shift, relative to the rest wavelength
of H$\alpha$, along the horizontal.  The colors are for brightness, as
indicated at far right.  The dynamic spectrum shows two orbital phases,
with the second being a repeat of the first, for ease of viewing.
The lower panel gives an average spectrum for all the data.

The extension to large blueshift velocities well outside of H$\alpha$
is shown because it captures an absorption feature from the atmosphere
of the Loser star, shown as a black sinusoid.  The full amplitude is
approximately 400 km~s$^{-1}$, twice the orbital speed of the Loser.

The H$\alpha$ profile is complex.  A FWHM is difficult to define,
because the line shape is asymmetric and because the peak can vary
significantly in value and velocity shift.  Based on the average
spectrum, the HWHM is around
225 km~s$^{-1}$, with the line wings typically extending out to 
$\pm 400$ km~s$^{-1}$.

Peak brightness in H$\alpha$ occurs around $\epsilon \sim 0$, which
is when the Loser, which is the brighter component, is in eclipse.
This certainly causes continuum-normalized line emission to increase
because the continuum level for the binary is minimzed.  One might
expect a similar brightening of H$\alpha$ during secondary minimum
around $\epsilon \sim 0.5$.  Using the same dataset as this paper,
\cite{2008JSARA...2...71G} showed that the equivalent
width of H$\alpha$ is anticorrelated with the observed photometric
light curve, and that the total line flux of H$\alpha$ is roughly
constant with phase, albeit with significant scatter.  The significant
brightening in the dynamic spectrum is approximately confined to
$\Delta \epsilon \approx \pm 0.1$ in phase, which corresponds well
to when the Loser is eclipsed by the disk, given the adopted
geometrical parameters in Table~\ref{tab1}.

One striking feature revealed by the dynamic spectrum is the blueshifted
absorption.  The feature appears somewhat
prominently between phases of 0.1 and 0.4, then again less prominently
from about 0.75 to 0.9.  The feature has a width of about 50 km~s$^{-1}$.
For the lower phase interval, the feature appears to drift from
about $-150$ to $-50$ km~s$^{-1}$.  For the higher phase interval,
it drifts perhaps from $-50$ to $-100$ km~s$^{-1}$, although it is
unclear whether any drift occurs, owing to the weak level of
absorption.  Key is that the feature never moves to positive velocity
shifts.

This component is clearly not ascribable to either the disk or jet,
based on our qualitative modeling.  While line asymmetry can result
in the emission lines formed in the disk owing to eclipse by the
Loser (c.f., Fig.~\ref{fig:diskphase}), the effect drifts from
blueshifts to redshifts (because the isovelocity zones are left-right
symmetric), and is localized around $\epsilon \sim 0.5$, exactly
where the observed component disappears.  By contrast the line
emission from the jet is always symmetric about the peak, even in
eclipse, because the isovelocity zones are front-back symmetric.
It is difficult to imagine how a detailed radiative transfer
calculation involving both the disk and the jet plus the eclipse
could produce the observed feature -- its persistent blueshift and
apparent drift -- even if the disk and jet were non-axisymmetric.

There are components in the $\beta$ Lyr system that have not been
considered in our modeling, such as a circumbinary envelope, a
hot spot, and the accretion stream.  We consider each of these in
relation to the drifting blueshifted feature in the dynamic spectrum
of H$\alpha$.

\begin{figure}[t]
\centering
\includegraphics[width=0.95\columnwidth]{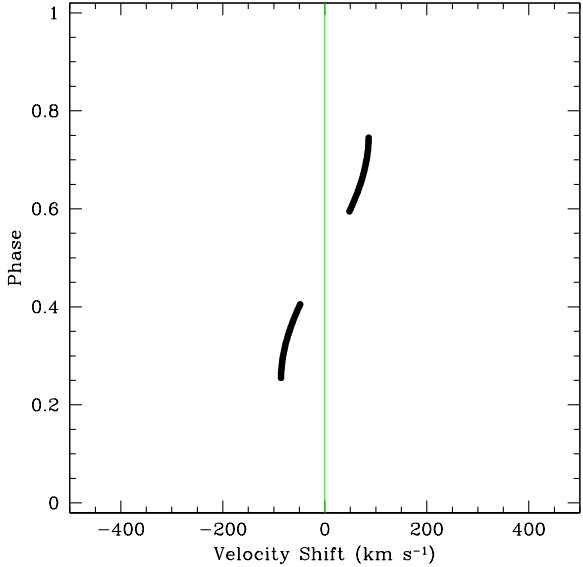}
\caption{Displayed is the path of a ``spot'' on
the disk (in emission
or absorption) in velocity shift with phase.  The spot
is not seen at early or late phases when on the far side of
the disk.  The gap in the middle is when the spot is
eclipsed by the Loser.  See text for further details.}
\label{fig:spot} 
\end{figure}

\begin{itemize}

\item {\em Circumbinary Envelope:}  A circumbinary envelope is
likely to be a torus-like region residing outside of the binary as
a whole.  If rotating, its rotation must be slow.  Treating the
binary as a point source of mass with $16M_\odot$, the distance at
which a rotation speed of 100 km~s$^{-1}$, a value consistent with
the feature, is achieved would be about $290R_\odot$, or roughly
$5.2\times$ the binary separation.  There are several challenges
for a circumbinary envelope explanation of the drifting feature.
How would the feature avoid appearing at redshifted velocities?
And why would it not be present for certain phase ranges?  

If the circumbinary torus were in expansion, then the feature
might be explainable as a low-density segment of the torus.
But then what is the origin for the non-axisymmetry?  The
binary itself is the obvious choice.  However, it seems challenging
for any stationary structure (i.e., in the rotating frame of the binary)
to avoid producing a feature at redshift velocities in the observer frame,
as the source of asymmetry (e.g., if mass loss were to occur
via the L2 and L3
points) would track with the binary orbit.

\item {\em Hot Spot:}  Many authors suggest the presence of a hot spot
where the accretion stream from the Loser lands at the accretion disk
for the Gainer \citep[e.g.,][]{2012ApJ...750...59L}.  
Consider a hot spot that is on the outer-facing rim of the vertically
extended accretion disk, as a zone of limited arc along the rim.  The spot
is stationary in the rotating frame.  For an edge-on observer, the spot is
not viewable around $\epsilon \sim 0$ because of occultation by the disk,
nor around $\epsilon \sim 0.5$ owing to eclipse by the Loser star.

A hot spot at the disk rim, as seen edge-on, would follow a velocity shift
given by

\begin{equation}
v_{\rm z} = -\Omega\,(R_D-\varpi_{\rm b1})\,\sin(2\pi\,\epsilon).
\end{equation}

\noindent The co-rotation speed at the disk rim, from the barycenter,
is $\Omega (R_D-r_{\rm b1}) \approx 80$ km s$^{-1}$.  The Keplerian
rotation at the rim is about the same.  The overall scale of these speeds
at the hot spot is commensurate with the observed maximum blueshift value
for the feature in H$\alpha$.
Figure~\ref{fig:spot} illustrates how such a feature would migrate through
a dynamic spectrum.  Here the spot is eclipsed by the disk for orbital
phases of 0.0-0.25, and then again for 0.75-1.0.  It is also eclipsed
by the Loser star at phase 0.5.  For illustration purposes the spot is
simply assumed to be along the line-of-centers between the two stars.
A hot spot produces a feature appearing at both blueshifted and redshifted
velocities, which is not observed.

For the mass transfer, one can imagine that the hot spot leads the Loser.
This would mean that the spot appears before $\epsilon = 0.25$, and it
would be eclipsed before $\epsilon = 0.75$.  Indeed, the geometry for
a spot that leads the Loser in phase could allow for an eclipse that
would largely block the spot by the Loser when 
at redshifted velocities.  The challenge is that the spot would
then be seen around $\epsilon = 0.5$, which is not observed.  Further,
it seems almost impossible to understand how a spot could 
produce a blueshifted feature after $\epsilon = 0.5$.

\begin{figure}[t]
\centering
\includegraphics[width=0.95\columnwidth]{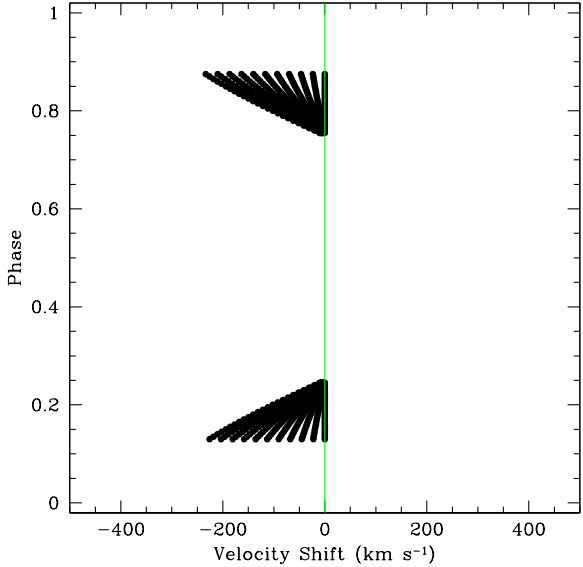}
\caption{Displayed is the path of material in the
accretion stream (in emission or absorption).  The gap
in the middle is when the origin of the accretion is
blocked by the Loser.  The gaps at early or late phases is
when the Loser is eclipsed by the disk.  The different lines are
for different flow speeds in the stream.  See text for
further details.}
\label{fig:stream} 
\end{figure}

\item {\em Accretion Stream:} For simplicity imagine
the accretion stream as a flow of material moving radially away from
the Loser, along the line-of-centers joining the stars, to arrive 
at the accretion disk.  This stream would be seen against the backdrop 
of the atmosphere of the Loser for phases below 0.25  and above 0.75.
It would not be seen between phases of 0.25-0.75, as it would be
blocked by the backside of the Loser.  In addition, the stream would
not be seen for phases of about $0.0-0.1$ and $0.9-1.0$, owing to occultation
of the stream by the disk.

But when the stream is observed, it would be moving from the Loser
toward the disk with a velocity component
toward the observer, and so would appear
strictly at blueshifted velocities, with $v_{\rm z} = -v_{\rm acc}\,
\cos(2\pi\,\epsilon)$, where $v_{\rm acc}$ is a free parameter for a
stream of constant speed.  Figure~\ref{fig:stream} shows how the kinematics of
this feature would appear in a dynamic spectrum.  The different lines
are for different values of $v_{\rm acc}$.  The lines replicate the
overall observed pattern in H$\alpha$:  the feature is blueshifted only;
is absent at appropriate phase intervals; moves from high blueshifts to
low ones as the Loser emerges from
eclipse; and then when appearing again, the drift reverses to move from
low to high blueshifts as the Loser moves toward ingress.

However, the observed feature is weaker at later phases than early phases.
There are some possibilities for explaining this.  The most plausible is
that, like the hot spot concept leading the Loser, the accretion stream 
is probably curved.  This would imply that the column of stream material
is more lined-up with the observer sightline at early phases than at lates
ones.  Such a detailed model is beyond the scope of this paper.

As plausible as this may seem, there is a severe challenge.  The model
so far neglects the contribution of the orbital motion for the kinematics
of the stream.  For early phases, the orbital motion adds a component of
velocity toward the observer, thereby increasing the blueshifted speed.
At late phases there is the addition of a redshifted component to the
net Doppler shift.  Figure~\ref{fig:streamorb} illustrates this effect.

\begin{figure}[t]
\centering
\includegraphics[width=0.95\columnwidth]{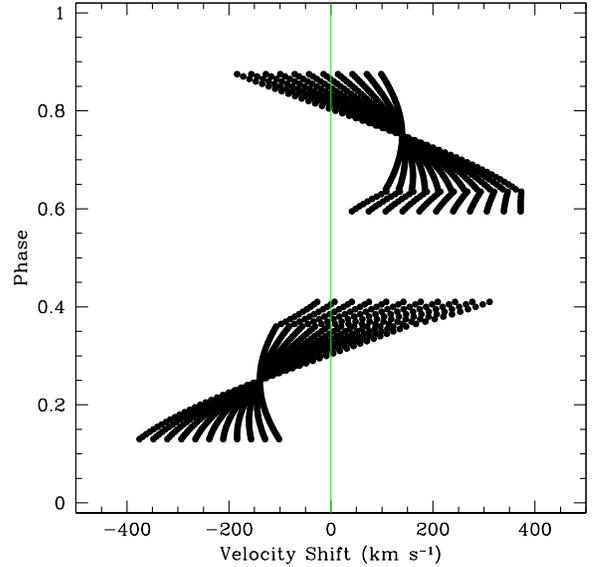}
\caption{Similar to Fig.~\ref{fig:stream}, but now with orbital
motion included.  See text for further explanation.}
\label{fig:streamorb} 
\end{figure}

Figure~\ref{fig:streamorb} is similar to Figure~\ref{fig:stream},
with different curves for difference choices of $v_{\rm acc}$.  A
constant orbital speed of about 140 km~s$^{-1}$ has been included,
with its phased-dependent velocity shift.  The selected orbital
speed is intermediate of the Loser's orbital speed of 180 km~s$^{-1}$
and that of the disk's rim at 80 km~s$^{-1}$.  In addition, we
allowed for the stream to be seen against the backdrop of the
disk, as not all of the stream is eclipsed by the Loser between
phases of 0.25-0.50.

The end result is that while the blueshifted portion of the curves
can have approximately the right speeds at the correct phases, considerations
of orbital motion result in the feature appearing at redshifts as well.
If $v_{\rm acc}$ is small, the early phases can approximate the
observations; however, at later phases the feature is entirely at
redshifts.  If $v_{\rm acc}$ is large, the feature still appears at
redshifts for late phases, but also at low blueshifted speeds.  However,
high blueshift speeds are achieved at early phases, contrary to
observations.  Further detailed modeling is needed to determine if
these shortcomings of the model can be overcome.  Allowing
for curvature of the accretion stream and acceleration of the accretion
flow would certainly greatly alter the kinematical signature
in a plot like Figure~\ref{fig:streamorb}.

\end{itemize}

\section{Conclusions}	\label{sec:conc}

Further dedicated monitoring of the H$\alpha$ emission line in
$\beta$~Lyr is needed to confirm suggestions that the well-known
persistent blueshifted absorption feature at low velocity shifts
in the line does indeed (a) come and go in certain phase intervals,
and (b) drifts in velocity.  Hints of such effects have appeared
in the literature.  \cite{1993PASP..105..426H} also reported on a
dynamic spectrum from a different dataset and claimed to observe an
``S-wave'' feature of relative variation in the line profile with
orbital phase.  However, their dataset had 1/3 as many spectra as
this report, and their phase binning was more coarse.  Using a
subset of the archival data from Ritter, \cite{2009ASPC..404..297A}
noted that the emission line can become single-peaked, indicating
that the blueshifted feature does not exist for all phases.

Based on the dataset described here, simplistic models were developed
for emission lines formed in the disk alone or in the jet alone.
Separately, these cannot produce the characteristic asymmetric
double-peaked profiles.  Evaluating the detailed radiative transfer
for a model involving both a disk and jet is unlikely to help.
Using basic kinematical considerations, neither a circumbinary component
(e.g., a ``torus'' in outflow or rotation) nor a hot spot seems
capable of explaining the appearance and disappearance of the feature
with phase, nor its migration in velocity shift.  An accretion
stream seems more promising, but also faces challenges.  
If new observations were to confirm the behavior of the feature 
as claimed here, the phase, velocity drift, and velocity width
information is likely probing dense, asymmetric flow components
in the system.  Perhaps the magnetic field claimed to exist in the 
system, yet rather poorly understood, could prove relevant
\cite[e.g.,][]{2018CoSka..48..300S}.

\acknowledgments

The authors thank an anonymous referee for several helpful comments to
improve this manuscript.  The authors also gratefully acknowledge that
archival data used for this paper and maintained by the University of
Toledo Ritter Observatory is supported by the National Science Foundation
Program for Research and Education with Small Telescopes (PREST).
This research has made use of the SIMBAD database, operated at CDS,
Strasbourg, France.  This research has made use of NASA's Astrophysics
Data System.

\appendix

\section*{Coordinate Transformations}	\label{app:coords}

The coordinate transformation between the observer system and
that of the barycenter is given by the following sets of
expressions, first for Cartesian coordinates:

\begin{equation}
\left(
\begin{array}{c}
x_{\rm b} \\
y_{\rm b} \\
z_{\rm b} \\
\end{array}
\right) = 
\left( \begin{array}{ccc}
\cos i & 0 & \sin i \\
0 & 1 & 0 \\
-\sin i & 0 & \cos i \end{array} \right)
\;
\left(
\begin{array}{c}
x \\
y \\
z \\
\end{array}
\right) ,
\end{equation}

\noindent and for spherical coordinates:

\begin{eqnarray}
\cos \chi & = & \cos\vartheta_{\rm b}\cos i + 
	\sin\vartheta_{\rm b}\sin i\cos\varphi_{\rm b},  \\
\cos \theta_{\rm b} & = & \cos i \cos \chi + \sin i\sin \chi\cos \alpha,  \\
\sin \chi \sin \alpha & = & - \sin \vartheta_{\rm b}\sin \varphi_{\rm b},  \\
\tan \alpha & = & - \frac{\sin \vartheta_{\rm b}\sin \varphi_{\rm b}}
	{\cos\vartheta_{\rm b}\sin i-\cos i\sin \vartheta_{\rm b}\cos 
	\varphi_{\rm b}} .
\end{eqnarray}

The transformation from the barycenter to the coordinates of the Gainer
is given by:

\begin{eqnarray}
x_1 & = &  x_{\rm b}-\varpi_{\rm b1}, \\
y_1 & = &  y_{\rm b}, \\
z_1 & = &  z_{\rm b},
\end{eqnarray}

\noindent and

\begin{eqnarray}
r_1^2 & = & x_1^2 + y_1^2 + z_1^2 \\
\cos \vartheta_1 & = & z_1/r_1 \\
\varpi_1^2 & = & x_1^2 + y_1^2 \\
\tan \phi_1 & = & y_1/x_1.
\end{eqnarray}

The transformation from the barycenter to that of the Loser is quite similar,
with the only difference being in the $x$-coordinate:

\begin{eqnarray}
x_2 & = &  x_{\rm b}+\varpi_{\rm b2}, \\
y_2 & = &  y_{\rm b}, \\
z_2 & = &  z_{\rm b},
\end{eqnarray}

\noindent and

\begin{eqnarray}
r_2^2 & = & x_2^2 + y_2^2 + z_2^2 \\
\cos \vartheta_2 & = & z_2/r_2 \\
\varpi_2^2 & = & x_2^2 + y_2^2 \\
\tan \phi_2 & = & y_2/x_2.
\end{eqnarray}

The coordinates for the jet system are defined in terms of the
Gainer coordinates.  The $+x$-axis for the jet coordinates is along
the outward cylindrical radial to the jet from the Gainer,
$+\hat{\varpi}_{\rm 1J}$.  The $z$-axis for the jet is parallel to
that of the Gainer, and so the $y$-axis for the jet is defined
by $\hat{z}\times\hat{x}$.
If the vertex of the jet is oriented in the disk plane at azimuth
$\varphi_{\rm 1J}$, then the Cartesian coordinate transformation
is given by:

\begin{equation}
\left(
\begin{array}{c}
x_1 \\
y_1 \\
z_1 \\
\end{array}
\right) =
\left( \begin{array}{ccc}
\cos \varphi_{\rm 1J} & \sin \varphi_{\rm 1J} & 0\\
-\sin \varphi_{\rm 1J} & \cos \varphi_{\rm 1J} & 0 \\
0 & 0 & 1 \end{array} \right)
\;
\left(
\begin{array}{c}
x_J+\varpi_{\rm 1J} \\
y_J \\
z_J \\
\end{array}
\right) ,
\end{equation}

\bibliography{blyr}

\end{document}